\documentclass[prl,aps,twocolumn, floatfix, superscriptaddress,showpacs]{revtex4}
\usepackage{amsmath,bm,graphicx}
\usepackage{amssymb}
\usepackage{amsfonts}
\usepackage{graphicx}
\usepackage{hyperref}
\usepackage{latexsym}

\def\be{\begin{equation}}
\def\ee{\end{equation}}
\def\bea{\begin{eqnarray}}
\def\eea{\end{eqnarray}}

\draft

\begin{document}
\title{Spontaneous oscillations of crosslinked treadmilling microtubules}
\title{Instrinsic oscillations of crosslinked treadmilling microtubules}
\title{Instrinsic oscillations of treadmilling microtubules in a motor bath}

\author{Sudipto Muhuri}
\affiliation{Departament de F\'{\i}sica Fonamental, Universitat de Barcelona, Carrer Mart\'{\i} i Franqu\`es 1, 08028 Barcelona, Spain}
\affiliation{Institute of Physics, Sachivalaya Marg, Bhubaneswar 751005, India}

\author{Ignacio Pagonabarraga}
\affiliation{Departament de F\'{\i}sica Fonamental, Universitat de Barcelona, Carrer Mart\'{\i} i Franqu\`es 1, 08028 Barcelona, Spain}

\author{Jaume Casademunt}
\affiliation{Departament d'Estructura i Constituents de la Mat\`eria, Universitat de Barcelona, Carrer Mart\'{\i} i Franqu\`es 1, 08028 Barcelona, Spain}


\begin{abstract} 
We analyse the dynamics of  overlapping  antiparallel treadmilling microtubules in the presence of crosslinking processive motor proteins that counterbalance an external force.
We show that coupling the force-dependent velocity of motors and the kinetics of motor exchange with a bath in the presence of treadmilling leads generically to oscillatory behavior.  In addition we show that coupling the polymerization kinetics to the external force through the  kinetics of the crosslinking motors can stabilize the oscillatory instability into finite-amplitude nonlinear oscillations and may lead to other scenarios, including bistability. 

\end{abstract}

\maketitle

The collective action of motor proteins plays a fundamental role in biological self-organization \cite{howard_book,Guerin2010}.
Spontaneous mechanical oscillations in biological systems where groups of motors are elastically coupled to their environment constitute a generic phenomenon~\cite{placais}.  Examples of such oscillations can be found, {\sl e.g.}  in the coupling of the mitotic spindle to the cell environment~\cite{julicher_prl, spindle_oscillations} or  the  mechanical oscillations in muscle fibers~\cite{kruse_njp}; the  oscillatory  behavior of motor-filament complexes has also been invoked to play a crucial role in mitosis~\cite{Campas_Sens}. Two main simplifications are usually considered in most studies of collective action of motors coupled to filaments, such as in gliding assays or other in vitro experiments~\cite{placais}: (i) the treadmilling dynamics of the filaments is eliminated; (ii) motors are fixed onto a substrate. In  biological situations, however, it is common that filaments are continuously treadmilling and motors can freely be exchanged with a motor bath in solution \cite{miyamoto,mitchison}. A paradigmatic example are the interpolar microtubules of the mitotic spindle, which undergo the so-called  poleward flux \cite{miyamoto}.

In this letter we show that force-dependent velocity of processive motors crosslinking treadmilling filaments against an external constant force, leads generically to intrinsic oscillations, which may be either damped or amplified.  We also show that, in the latter case, the oscillatory instability can be stabilized into nonlinear oscillations by the coupling of kinetics of  motor-exchange with microtubule polymerization.

To elucidate the generic character of the phenomenon, we focus on a minimal arrangement, inspired by the interpolar microtubules in the mitotic spindle, consisting of a symmetric pair of polymerizing antiparallel microtubules (MTs). These are crosslinked by (processive) tetrameric kinesins (Eg5) in a central region of overlap of size $\ell$, typically much smaller than the MT length \cite{miyamoto}.  The MTs overlap in their plus ends, the extremum where they polymerize as they are being pulled by the motors. By symmetry,  the crosslinking motors  remain in the overlap region while they slide the two MTs in opposite directions towards their respective minus ends (the so-called poleward flux of spindle MTs). At the same time, the crosslinking motors balance an external inward force $F$ that is assumed constant.  

In Ref. \cite{epl_otger} it was shown that under some plausible assumptions, the dynamics of the overlap region can be decoupled from that of the MTs outside, which enters through boundary conditions. Then, the motor kinetics in the overlap region may be described by simple rate equations 
for two populations of motors in the overlap region, $n_c$ designating the average number of motors crosslinked to both MTs, 
and $n_b$ those bound to either one of them. 
The kinetic equations take the simple form 
\begin{eqnarray}
\frac{dn_c}{dt} &=& - 2 k_{u}(n_c) n_{c} + k_{b}n_{b}
\label{eq:evolution_nn1} \\
\frac{dn_b}{dt} &=& 2k_{u}(n_c) n_{c} - (k_{b}+k_{u}^{0} )n_{b} + k_{b}^{3D}\rho_{3D}\ell, 
\label{eq:evolution_nn2}
\end{eqnarray}
where $k_b$ and $k_u^0$ are respectively the binding and unbinding rates of the motor domains for motors that are not under load. Only crosslinked motors exert forces and we assume a force-dependent kinetics of the standard form $k_{u} = k_{u}^{0}\exp(f_{m}b/k_{B}T)$, where $b$ is a length in the nanometer scale characterizing the activation process, $k_{B}T$ is the thermal energy, and the force per motor is taken as $f_m=F/n_c$. This results in a strongly nonlinear dependence of $k_u$ on $n_c$. We keep this force-dependent kinetics to remain quantitatively as realistic as possible although 
this dependence is not necessary to explain the qualitative picture of emergence of oscillatory behavior. 
The exchange with the motor bath is described by the balance between the last two terms in Eq.(\ref{eq:evolution_nn2}) \footnote{ In Eq.(\ref{eq:evolution_nn1}) a term of incoming bound motors from the non-overlaping region such as that considered in Ref.  \cite{epl_otger} could be introduced, given the processivity of motors. For simplicity, and since such effect is relatively small we omit this term here.}.
The motor intake from the bath is controlled by the rate $k_{b}^{3D}$ and is proportional to the motor density in solution $\rho_{3D}$, which together with $F$ define the two experimental control parameters of the problem.  In general, the overlap length, $\ell$, will evolve dynamically as a result of the balance  between the overall polymerization,  $V_p$, and sliding,  $V_s$,  velocities
\begin{equation}
\frac{d\ell}{dt} = 2 (V_p  - V_s) = 2 V_p (n_c) - 2V_{0}\left(1 - \frac{F}{n_{c}f_{s}} \right) 
\label{eq:evolution_nn3}
\end{equation}
thus relaxing a strong assumption of Ref.  \cite{epl_otger}. For
simplicity and consistently with experimental evidence, we have assumed a linear decrease of motor velocity relative to the MT with the force per motor $f_m=F/n_c$, and where $f_s$ is the single-motor stall force. The dependence of $V_p(n_c)$ expresses the coupling between motor dynamics and polymerization. 
Since the emergence of the oscillatory behaviour itself does not depend on this coupling,   
for the first part of the analysis we will take $V_p=const$. Later on we will discuss how this additional coupling may actually stabilize nonlinear oscillations.

Eqs. (\ref{eq:evolution_nn1})-(\ref{eq:evolution_nn3}) provides the essential description of the motor-microtubule complex. They can be expressed in dimensionless form in terms of  the characteristic detachment force, $k_BT/b$, the processivity length, $V_0/k_u^0$, and the detachment rate, $k_u^0$. Accordingly, we will use the dimensionless variables  $\tilde F = Fb/K_{B}T$,  $\tilde f = f_{s}b/K_{B}T$, $\tilde n_{c} = n_{c}/\tilde F$, $\tilde n_{b} = n_{b}/\tilde F$, $\tilde \ell = \ell \tilde{f} k_u^0/ V_0$ and  $\tau = t k_{u}^{0}$. We define $\Delta = \frac{\rho_{3d}k_{b}^{3d} V_0 k_{b}}{\tilde F \tilde f (k_{u}^{0})^3}$, a ratio of the strength of motor influx  and outflux  in the overlap region, as a dimensionless parameter that expresses the balance between the two external control parameters that couple the system with the environment. In turn, $\gamma  = {k_b}/{k_{u}^{0}}$ measures the asymmetry in motor attachment/detachment at vanishing load. 
%
The relevant parameters become then $\Delta$, $\gamma$ and $g = \tilde{f}(1 - V_p/V_0)$.  Eqs. (\ref{eq:evolution_nn1})-(\ref{eq:evolution_nn3})  have a stationary solution of the form
\begin{equation}
\tilde n_{c}^{f} = \frac{1}{g}, \;\;\;\;
\tilde n_{b}^{f} = \frac{1}{g\gamma }\exp[g], \;\;\;\;
\tilde \ell^{f} = \frac{1}{g\Delta }\exp[g] -2g,
\label{eq:steady}
\end{equation}
which holds for 
$g>0$ ($V_p < V_0$).
 Physically meaningful configurations are further restricted to the regime   $\Delta < \frac{\exp g}{2g^{2}}$ to ensure $\ell \geq 0$.  
 
\begin{figure}
\begin{center}
\includegraphics[width=6.0 cm,angle=-90]{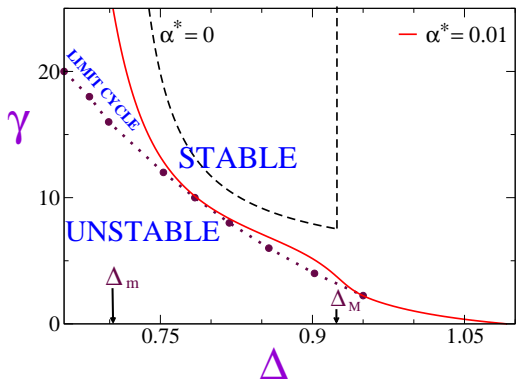}
\end{center}
\caption{Stability diagram of MT array: The stability line for $\alpha^{*} = 0$ (dashed curve) is derived from Eq.(\ref{eq:curve_stability}), where no limit cycle behaviour is observed. For $\alpha^{*} = 0$, the MT array is unstable below $\Delta_m=6 e^{7/4}/7^2$ and above $\Delta_M=e^2/2^3$ (as MT arrays are unphysical with $l < 0$). The area enclosed within the dashed curve is the region of stable arrays for  $\alpha^{*} = 0$. Stability  region is enhanced for inhibitory motors with $\alpha^{*} = 0.01$. The smooth curve (red online) depicts the stability line for this case. Between the smooth and dotted curves (maroon online), the arrays are nonlinearly stabilized and limit cycles are observed. Here $g = \frac{7}{4}$.}
\label{stability}
\end{figure}

The linear stability of the fixed point is determined by the eigenvalue  equation
\begin{equation}
\lambda^{3} + (1 + \gamma + G) \lambda^{2} +  (2g^{2}\Delta - G) \lambda + 2g^{2}\Delta = 0.
\end{equation}
where we introduce $G \equiv (g-1)\exp(g)$ for simplicity.
The  conditions for the existence of three real roots, and the corresponding regions of stability can be found explicitly, and will not be discussed here. We focus on the more interesting  case where two complex conjugate modes appear, 
\begin{eqnarray}
\lambda_1= - 1- \gamma +G \\
\lambda_{2,3} = \lambda_0 \pm i \theta,
\end{eqnarray} 
where $\lambda_0$ is real and is known explicitly in terms of the model parameters, 
$\lambda_1$ is always negative, and $\theta = \sqrt{2g^{2}\Delta - G}$. The stability region corresponds to $\lambda_0<0$, while the stability boundary $\lambda_0=0$ is defined by 
\begin{equation}
\frac{\gamma}{G} = 1 + \frac{1}{2g^{2}\Delta - G}.
\label{eq:curve_stability}
\end{equation}
where stability also  requires $2 g^2\Delta-G  > 0$, imposing a lower bound on $\Delta$. This constraint, together with $\ell  \geq 0$  does not allow for stable, physical steady state configurations when  $ g >2$. Since $\gamma$ and $\Delta$ are positive parameters, the steady antiparallel arrays are intrinsically stable for $g<1$, regardless of $\Delta$. Only  in the regime $1<g<2$ the motor/MT complex stability is controlled by  $\Delta$.  
 Fig.~\ref{stability} displays  the stability curve, as a dashed line,  for $g=7/4$ where stable antiparallel structures exist in a bound region of parameters, $\Delta_m \leq \Delta \leq \Delta_M$. The instability associated to the change of sign of $\lambda_0$ corresponds to a Hopf bifurcation, giving rise to damped or growing oscillations at a finite frequency respectively at the stable and unstable sides of the stability boundary. The characteristic dimensionless frequency $\theta = Im (\lambda_2)$ is given in full units by 
$\omega_B=k_u^0 \sqrt{2g^{2}\Delta - G}$ and  depends both on kinetic parameters and  the environment through $\Delta$; 
the oscillatory behavior persists outside the neighbourhood of the instability, with a weak dependence of the frequency on nonlinearities.  

To gain further analytical insight, it is useful to focus on the region near the instability threshold, where $|\lambda_1| \gg |\lambda_0|$. We can then write $\lambda_0 \simeq G-\gamma$ and $\lambda_1 \simeq -1 + \lambda_0$ and exploit the separation of time scales between the (slow) oscillatory modes and the (fast) relaxation mode. This is a rigorous two-dimensional reduction of the problem (to the so-called center manifold) which can be extended to the nonlinear level.
The adiabatic elimination of the fast mode $\delta \phi_1\sim G \delta \tilde{n}_c + \gamma \delta \tilde{n}_b  - \theta^2  \delta \tilde{\ell}/2g^2$, which decays as $\delta \phi_1 \sim \exp{(-\lambda_1 t)}$, 
implies that it will be slaved to the oscillatory modes.
 It is therefore legitimate to take $\delta \dot{\phi_1} \simeq 0$ which introduces a constraint on the variations of the three variables. Hence,  at linear level there is an approximately conserved quantity of the form
\begin{equation}
\frac{1}{\tilde{\ell}} \left( G \tilde{n}_c + \gamma \tilde{n}_b \right)  \simeq \frac{\theta^2}{2g^2}.
\end{equation}
This implies that the overlap length, $\tilde{\ell}$, reacts to keep constant an effective motor density that weights the two types of motors differently; only when $\Delta \gg 1$ this weighted density reduces to the total motor density. Consequently, as long as $|\lambda_1| \gg |\lambda_0| \simeq |G-\gamma| $, the dynamics of $\delta n_c$ at linear level is that of a harmonic oscillator 
\begin{equation}
\delta \ddot{\tilde{n}}_c - \lambda_0 \delta \dot{\tilde{n}}_c + \theta^2 \delta \tilde{n}_c =0. 
\end{equation}
where the oscillations of frequency $\theta$ are exponentially damped or amplified depending on the sign of $\lambda_0$.
The other two variables have the same behavior with different amplitudes and phases which follow from the linear relation $\delta \dot{\tilde \ell}=-2g^2 \delta \tilde{n}_c$. 

It is important to remark that the overlap length $\ell$ must be coupled to the motor dynamics in two ways in order to produce oscillations: (i) through the motor exchange in the bath with a motor intake proportional to $\ell$; and (ii) through the decreasing velocity-force curve of the motors, implying that  $\delta \dot{\ell}  \sim -\delta n_c$. The existence of oscillatory behavior is independent of the detailed dependence of motor kinetics and motor velocity on the applied force (as long as it is monotonously decreasing). The kinetic exchange between $n_b$ and $n_c$ is nevertheless essential to control the existence of a nontrivial stationary point (with finite values of all variables) and its stability, in particular allowing for unstable growth of the oscillations.  A nonlinear analysis is necessary in that case, and model details may come into play.

For  constant $V_p$, the amplitude of the oscillations is not saturated by nonlinearities leading  eventually to array disassembly. The coupling to the external environment or additional dissipation provides a means to 
 stabilize the filament-motor assemblies. Remarkably, simple assumptions on the coupling between motor kinetics and  MT polymerization 
provide possible mechanisms of nonlinear stabilization of the oscillations in some  regimes, introducing new dynamical scenarios. As an illustration we may consider $V_p(n_c)= \left[C_{p} (n_c)- C_{d}(n_c)\right]\epsilon$ where $C_p$ and $C_d$ are the polymerization and depolymerization rates at the plus end, with $\epsilon$ the length increase per added monomer. If the coupling is weak, one can linearize the dependence of motor concentration in the polymerization rates, rendering  the detailed functional form of the coupling irrelevant. Accordingly, one can generically write $C_{p(d)}\simeq C_{p (dp)}(1-\alpha_{p(dp)} n_c/\ell)$. In this weak coupling limit, it is enough to consider a single parameter, $\alpha_0$, the coupling strength, which characterizes the relative magnitude of the polymerization and depolymerization rates~\footnote{ In-vitro experimental evidence shows that processive plus-end motors enhance MT depolymerization in budding yeasts~\cite{tolic, howard1}. Nonetheless, the interactions of molecular motors with MT in other situations, and the possibility that other molecules may modify the effective motor/MT interactions leaves this issue generically open. In our approach all these details are effectively included in the phenomenological parameter $\alpha$. }; 
 motors  promote  depolymerization for $\alpha_0 < 0$ and inhibit it for $\alpha_0 > 0$, and their effects can be quantified in terms of the 
%
dimensionless parameters $\alpha = {\alpha_{0}\tilde F \tilde f}/{l_p}$, $\Gamma_1 = {\epsilon C_{p}}/({k_{u}^{0} l_p})$ and $\Gamma_2  = {\epsilon C_{dp}}/({k_{u}^{0} l_p})$, giving now  $g= \tilde{f}[1 -(\Gamma_{1} - \Gamma_{2})]$.
%
Eq.~(\ref{eq:evolution_nn3}), takes now the dimensionless form
\begin{equation}
\frac{d\tilde \ell}{d\tau} = -2g + \frac{2}{\tilde n_c} + \frac{2\alpha^{*}\tilde n_{c}}{\tilde \ell}
\end{equation}
where $\alpha^{*} = \tilde{f}\Gamma_2\alpha$. The steady solution then reads,
\begin{equation}
\tilde \ell^{f} = \frac{\alpha^{*}\tilde n_{c}}{g - 1/\tilde n_{c}}, \;\;\;\;
\tilde \ell^{f} =\frac{\tilde n_{c}}{\Delta}\exp(\frac{1}{\tilde n_{c}}) - \frac{2}{\tilde n_{c}},
\label{eqfixedpoint}
\end{equation}
along with $\tilde n_{b} = \frac{\tilde n_{c}}{\gamma}\exp(\frac{1}{\tilde n_{c}})$.  For 
$\alpha^{*} \ll \tilde \ell^{f}g^{2}$, the  steady antiparallel array changes gradually with  $\tilde n_{c} = 1/g + \alpha^*/\tilde \ell^{f}g^{3} $.

%
The nonlinear dependence between $n_c$ and $\ell$ allows for physical solutions if $\tilde{n}_{c,m} > 1/g$, $\tilde{\ell}_m< 4 \alpha^*/g^2$ and $1/\Delta > \alpha/g$, which  imply  $g\geq 2$. As a result, qualitatively new steady arrangements appear,  unaccessible for 
$\alpha=0$ . For $g\geq 2$ two new  steady configurations are allowed, and the overlap  region has always a positive  length regardless of $\Delta$.
The numerical analysis shows that polymerization enhancement, $\alpha^*>0$, favors the array stability, as shown in Fig.~\ref{stability} for $\alpha^*=1/100$,  while the opposite holds if motors enhance MT depolymerization~\footnote{\rm When motors promote MT polymerization the restriction $g<2$ holds. In this case $\Delta_m$ and $\Delta_M$ approach as $\alpha$ decreases, reducing the parameter range where antiparallel  arrays are stable.}. 

\begin{figure}
\begin{center}
\includegraphics[width=6.0 cm,angle=-90]{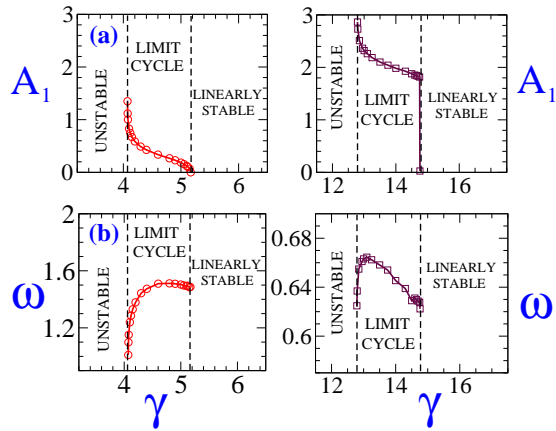}
\end{center}
\caption{(a) Variation of $A_{1}$,  the amplitude of the non-linear oscillations of the overlap region, $\tilde l$, with $\gamma$ for  $\alpha^{*}= 0.01, g = 1.75$.  Circles (red curve online) corresponds to $\Delta=0.9$, when the instability is supercritical and Squares (maroon curve online) for $\Delta =0.74$, when it is subcritical. b) Variation of the frequency of the limit cycles as a function of $\gamma$ for the corresponding points.}
\label{amplitude}
\end{figure}

For inhibitory couplings,  $\alpha>0$, linearly unstable arrays can be stabilized nonlinearly  close to the stability curve leading to  limit cycle oscillations.
 The amplitude of these nonlinear oscillations increases  when  moving into the unstable region, leading eventually to  array disassembly.
For $\alpha=1/100$,  Fig.~\ref{stability} shows the  region where nonlinear oscillations are sustained.  Fig.~\ref{amplitude}(a) identifies a regime, $\Delta>0.8$ where the amplitude vanishes as the square root of the distance to the linear stability threshold, analogous to a second order transition while for $\Delta<0.8$, the nonlinear stabilization has a finite amplitude from the outset, reminiscent of a first order transition, as shown in Fig.~\ref{amplitude}(b).The oscillation frequency always varies as we move away from the instability threshold, as depicted in Fig.~\ref{amplitude}. 
As one approaches the nonlinear stability threshold, decreasing $\gamma$, the frequencies decrease. We have never  observed nonlinearly stabilized arrays when motors promote MT depolymerization ($\alpha \leq 0$). 

\begin{figure}
\begin{center}
\includegraphics[width=6.0 cm,angle=-90]{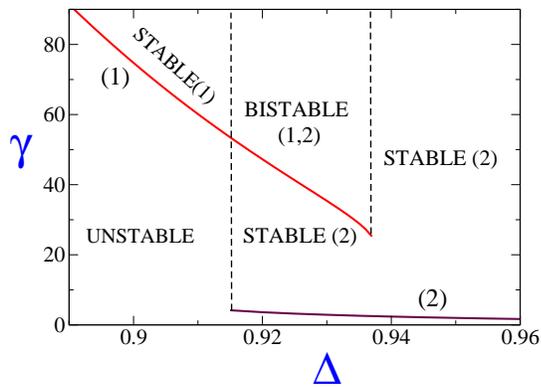}
\end{center}
\caption{Stability diagram of an antiparallel MT array for  $g = 2.7$ and $\alpha^* = 0.1$. At large $\gamma$ there is a region of intermediate $\Delta$ values where both a short (1) and a long (2)  MT arrays are stable, leading to bistability.}
\label{fig:bistable}
\end{figure}
The existence of three steady configurations for $g\geq 2$ when $\alpha > 0$ allows for new dynamic scenarios. 
As shown in Fig.~\ref{fig:bistable}, by varying $\Delta$ it is possible to find bistability.
The  coupling between the array nonlinear oscillations with different external  frequencies  in this bistable regime provides new scenarios to  control the motor-MT complexes stability and its sensitivity to  changes in the array environment.

We have shown that the self-organized, coupled dynamics of the overlapping region of antiparallel MT arrays gives rise to intrinsic oscillations due to the  localized motor kinetics in the overlap region and its coupling with MT polymerization. These oscillations may be relevant in spindle dynamics, where oscillations reported to date emerge from the interaction of the antiparallel array with the surrounding media through centrosomal microtubules~\cite{julicher_prl}, or the interaction of molecular motors  with  the chromosomes MT attach to~\cite{Campas_Sens}. The reported oscillations are generated at  relatively small scales, of the order of the few fractions of a micrometer which characterize the overlap between biofilaments, but the frequency range, controlled by the motor unbinding rates, is  typically of the order of  $s^{-1}$. The coexistence of different stable arrays also allows for bistability between antiparallel arrays, providing enhanced mechanical versatility of these structures to environmental changes.

\acknowledgments
Financial support from MICINN  (Spain) and Generalitat de Catalunya is acknowledged, by IP and SM under projects  FIS2008-04386 and 2009SGR-634, respectively, and by JC under projects FIS2010-
21924-C02-02 and  2009-SGR-014, respectively

\end{document}